# Astrophysics in Southern Africa

Patricia Whitelock

*National Astrophysics and Space Science Programme, Department of Mathematics and Applied Mathematics and Department of Astronomy, University of Cape Town, and South African Astronomical Observatory, P O Box 9, Observatory, 7935, South Africa. paw@saao.ac.za*

**Abstract.** The government of South Africa has identified astronomy as a field in which their country has a strategic advantage and is consequently investing very significantly in astronomical infrastructure. South Africa now operates a 10-m class optical telescope, the Southern African Large Telescope (SALT), and is one of two countries short listed to host the Square Kilometre Array (SKA), an ambitious international project to construct a radio telescope with a sensitivity one hundred times that of any existing telescope. The challenge now is to produce an indigenous community of users for these facilities, particularly from among the black population which was severely disadvantaged under the apartheid regime. In this paper I briefly describe the observing facilities in Southern Africa before going on to discuss the various collaborations that are allowing us to use astronomy as a tool for development, and at the same time to train a new generation of astronomers who will be well grounded in the science and linked to their colleagues internationally.

**Keywords:** Africa, Astronomy
**PACS:** 01.10.Hx, 01.40.-2, 01.75.+m, 01.78.+pe, 95.45.+i

## INTRODUCTION

A new science policy in South Africa was one of many changes that followed from the election of a democratic government in 1994. The policies of the previous regime had been dominated by the requirements of apartheid and its consequences, which included isolation from the international scientific community. The new government were driven by the clear need for poverty alleviation and a strong desire to overcome the many inequalities that were, and remain, legacies of apartheid. They had, from South Africa, the backing of the majority of the people, and from the international community a strong political will to participate in the reconstruction of the country, including its science policy and infrastructure. It is within this atmosphere that the developments described here are being made.

1994 saw the birth of a Department for Arts, Culture, Science and Technology within the government, which in 2002 split into two parts, the Department of Science and Technology (DST) being the one of interest to us here. It is within this department and its agencies, in particular the National Research Foundation (NRF), that the developments described below were formulated and taken forward. The exploitation of our resources and our strengths was fundamental to the new approach for tackling poverty, its causes and its consequences. As stated in the 2002 policy document from DST [1]: "Clearly, with limited resources our best chances of success will depend on our ability to focus on our potential strengths while staying well connected to international research", and "Scientific areas in which there is an obvious geographic advantage:
1. Astronomy
2. Human Palaeontology
3. Biodiversity
4. Antarctic Research"

Astronomy was chosen as a beneficiary of the new dispensation partly because of geographic advantages, latitude, longitude, climate, etc., partly because of the subject's proven ability to draw smart young people into the science arena and partly because of existing strengths in the field. This and related issues were discussed in more detail elsewhere [2].

In the following I briefly describe the facilities available to astronomers in Southern Africa, before discussing the primary postgraduate training programme which is aimed to ensure that these magnificent facilities will not be staffed entirely by scientists from outside the country.

## ASTRONOMY FACILITIES

South Africa runs or is a partner in the following facilities or proposed facilities. The level of participation in the new projects varies from that of lead partner and instigator (SALT) to that of a small contributor (HESS).

### Existing National Facilities

The South African Astronomical Observatory (SAAO) is the National Facility for optical and infrared astronomy in South Africa. Its headquarters are in Cape Town and its observing facilities at Sutherland in the Northern Cape. It has four common user telescopes with apertures ranging from 0.5 to 1.9-m as well as a 0.5-m automated telescope. SAAO is also host to a variety of small telescopes run on behalf of, or in collaboration with, Korea, Japan, Germany and the UK as well as a geodynamic observatory and various seismographs. The site is well regarded for its seismic stability as well as its dark and photometrically-stable skies. See www.saao.ac.za.

The Hartebeeshoek Radio Observatory (HartRAO), situated in Gauteng, is the National Facility for radio astronomy. It has a 26m dish built by NASA, but resurfaced locally to reach frequencies up to 22 GHz. HartRAO also has a satellite laser ranging facility and an IGS GPS network and engages in geodesy as well as in radio astronomy. Its position as the only such facility on the African plate makes it a key resource for very long baseline interferometry (VLBI) in collaboration with Europe, USA, Australia or even orbiting facilities. See www.hartrao.ac.za

## High Energy Stereoscopic System (HESS)

The HESS telescope array is situated in Namibia, near the Gamsberg mountain. It detects very high energy, 100 GeV, gamma-rays via their Cherenkov radiation and achieves a stereoscopic reconstruction of the position and spatial structure of the source of emission using four telescopes. It is an international, dominantly European, project. First light was in 2002 and all four telescopes were operational by the end of 2003. South African participation is via the University of the North West, but broader participation will be possible in HESS II. This is under construction following the great success of the first experiment in a variety of areas, including the detection of the Centre of our own Galaxy where results support the existence of a super-massive black hole. More information is available from:
www.mpi-hd.mpg.de/hfm/HESS/.

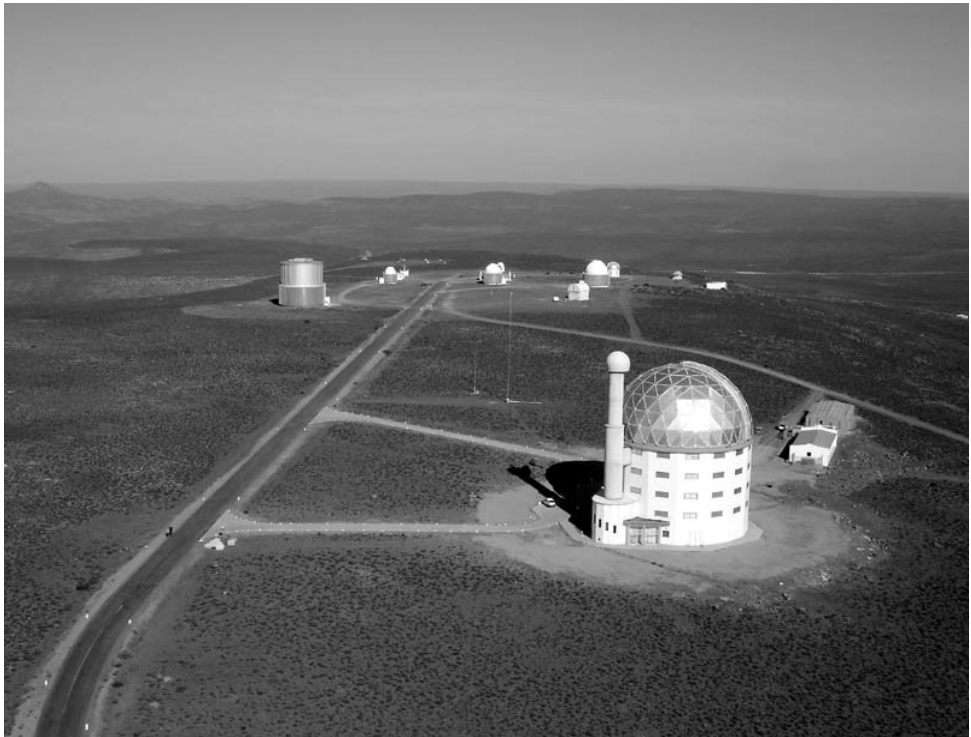

*Illustration 1: SALT and SAAO telescopes at Sutherland in South Africa, Copyright: Helicam and Atlantis Panel*

## Southern African Large Telescope (SALT)

SALT is a 10-m class optical telescope modelled on the Hobby-Eberly Telescope (HET) in Texas, although with a significant number of upgrades. It is an international collaboration, with approximately one third belonging to South Africa, one third to various USA universities and the rest to countries or organizations within Europe. It is the single largest optical telescope in the Southern hemisphere and was built at about 20% of the cost of a conventional telescope. It is sited at Sutherland in the Northern Cape Province at an altitude of 1800m.

SALT was opened by President Mbeki in 2005 and the first science from it was published the following year. The telescope is optimized for observations in the near-ultraviolet, down to the atmospheric cut off, and for high time resolution. Its main instrument is a low to medium resolution multi-object spectrograph, which also allows for spectra-polarimetry and Fabry-Perot imaging. More details can be found at: www.salt.ac.za.

## Karoo Array Telescope (MeerKAT) and Square Kilometre Array (SKA)

The SKA is a $2-billion project to build a radio telescope with a hundred times the sensitivity of any existing telescope. It is a truly international project and its feasibility will ultimately depend on new technology, hardware and IT, that has yet to be developed. The SKA will be capable of probing the very early universe and of providing new insight on the very first generation of stars and galaxies. Several countries bid to host SKA and in 2006 the two finalists were chosen: Australia and South Africa.

The MeerKAT (meer is Afrikaans for "more") has arisen out of South Africa's ambition to host the SKA and will be one of several science and technology "pathfinders" currently under construction around the globe. It is positioned in a radio quiet zone at the proposed site for the core of the SKA, and will form part of the SKA should South Africa be selected as the site. It will have 2-4% of the sensitivity of the full SKA (perhaps implemented as 80 15-m dishes), and will exceed the survey speed of all existing telescopes by orders of magnitude. A precondition for government investment in MeerKAT is the participation of international partners, and such partnerships are currently being established.

## ASTRONOMERS

Table 1 lists the various universities and other facilities within South Africa where astronomical research is done. The number of astronomers at each institution is given, including those nominally retired when they are still publishing. I have counted all PhD qualified scientists who could loosely be described as "astronomers", including several theoretical cosmologists who are certainly not direct users of the facilities, but who are an important component of a balanced research community. In the cases of some institutions, e.g. HartRAO/MeerKAT, Wits and UCT the "astronomers" are split between two or three different departments or locations.

The number of astronomers in South Africa has been growing slowly, there were fewer than 50 ten years ago. Nevertheless, it is very clear from this table that people and the facilities are not well matched. Furthermore the only critical mass of astronomers is in Cape Town (SAAO, UCT, UWC and part of MeerKAT).

Note, in particular, that only 8 of the 61 people listed above are black (note that in 2006 white people comprised only 9.2% of the South African population of 47.4 million) and less than 50% of the rest were born South African; 8 of the 60 are women. South Africa has a very poor record of producing South African astronomers and particularly so when it comes to black astronomers.

**TABLE 1. Numbers of Astronomers at various institutions in 2006**

| Institution | Number |
| --- | --- |
| SAAO | 25 |
| University of Cape Town (UCT) | 12 |
| HartRAO/MeerKAT | 5 |
| University of KwaZulu-Natal | 4 |
| University of the North West | 3 |
| University of the Witwatersrand (Wits) | 3 |
| University of South Africa | 3 |
| University of Johannesburg | 2 |
| University of the Western Cape (UWC) | 1 |
| University of the Free State | 1 |
| University of Zululand | 1 |
| Rhodes University | 1 |
| **TOTAL** | **61** |

# National Astrophysics and Space Science Programme (NASSP)
## *Background*

It became clear to the astronomy community in 2001, during the strategic planning exercise that eventually led to South Africa's participation in SKA, that the biggest threat to the future of astronomy on the subcontinent was the lack of South African astronomers and particularly the lack of black astronomers. Something had to be done or we would end up running superb international research facilities as technicians for our international partners, without the capacity to participate at the cutting edge. The long term political consequences of that would be disastrous. Astronomy is being supported at an extraordinary level for pure science with an expectation of delivery on many fronts. One natural expectation is that South Africans will fully participate in the exploitation of these facilities.

There are two critical points at which young people leave the road which could lead them to a PhD in astronomy. The first is at "matriculation", the final school-leaving exam, when fewer than 1% of black South Africans obtain a mathematics pass at the level that will allow them to study science or engineering at university. This is unquestionably a legacy of apartheid, which specifically excluded black people from studying maths and science, but it is also a sad fact that the situation has improved only slightly since 1994. The second loss point is after the first degree; very few students go on to postgraduate studies that will ultimately lead them to a PhD - the minimum qualification for a research career. This is recognized as a problem that goes way beyond astronomy; the country is not producing enough PhDs to replace its ageing academic population, never mind to provide for the knowledge economy that government strategy aims to nurture.

The initiatives that are attempting to deal with the first loss point mentioned above are outside the scope of this contribution, so I will only discuss the second one here.

In 2001 there were about 50 astronomers spread among the institutions listed above, and no single university had enough staff to run a proper postgraduate programme in the field. The community therefore decided to combine forces and create what we now call NASSP, with the primary aim of getting a viable number of BSc graduates to a point that they could start PhDs[1].

This was, and remains, very difficult, because of the way universities are financed in South Africa, which encourages them to compete rather than to collaborate and pushes, e.g. physics, departments to go to great lengths to prevent other universities or even other departments within the same university from "stealing" their students.

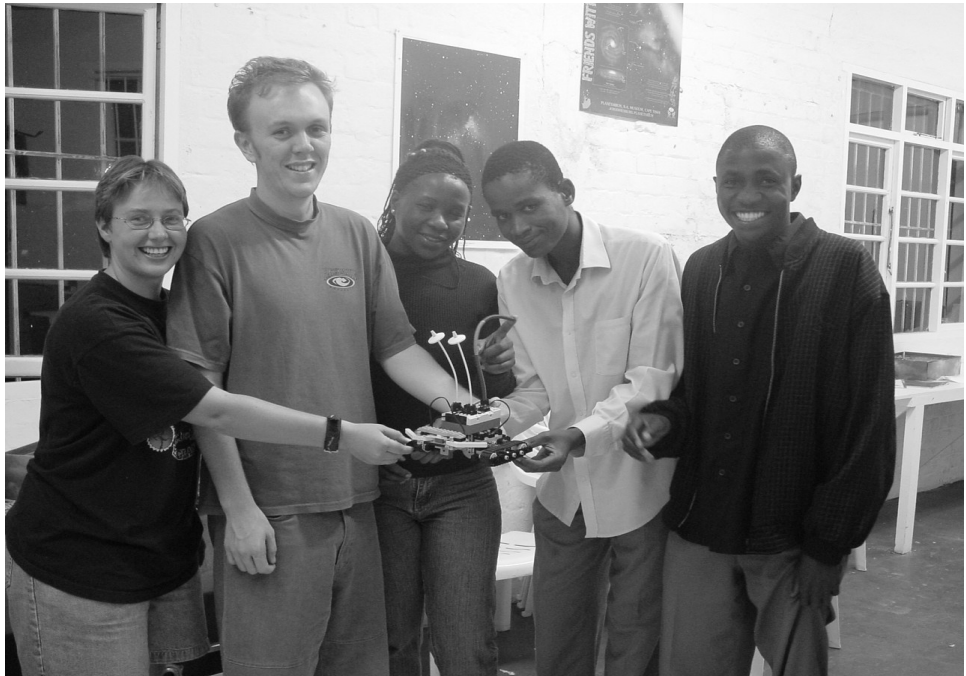

*Illustration 2: Students at the NASSP "preschool" learn to work as a team by programming a toy robot. Copyright: SAAO.*

---

[1] Within the South African system students must get at least a 1yr honours and usually also a 1 to 2 yr masters degree, before they qualify to start a PhD; the efficacy of this is debatable.

*NASSP as it now functions*

NASSP recruits students from all over South Africa, and elsewhere in Africa, to come to a single university and study together. Similarly the NASSP lecturing staff come from all over South Africa to the same university to teach intensively for periods of three or four weeks. The University of Cape Town (UCT) was the institute selected to host NASSP, because of its proximity to the large research community within SAAO who would also participate in the teaching. NASSP is based in the Mathematics and Applied Mathematics Department at UCT, because that is where Peter Dunsby, the current coordinator, is affiliated, but the Departments of Astronomy and Physics also participate in the teaching.

The detailed NASSP curriculum can be seen on the web site [www.star.ac.za](www.star.ac.za). For honours students it is a broad mixture of observational (optical and radio) and theoretical astronomy with an emphasis on skills, and it includes a research project. All of the courses are compulsory, as we aim to give those who become theorists some insight into observational astronomy and vice versa. At the masters level the students do course work for 6 months and can choose between the broad streams of extragalactic astronomy, stellar astronomy, cosmology and space physics. As the community grows and strengthens we aim to broaden the choices available at the master's level. After they have completed the course work the students choose their thesis topic and move to the institution of their supervisor for their research year during which they must write a mini-thesis. Their final qualification depends on both the exam results from the 6 months of course work and on the thesis; the degree is awarded by the university to which the supervisor is affiliated. Once a student has started on their thesis it is possible for them to upgrade and register for a PhD without first completing the masters; this is by the mutual agreement of all parties. Several NASSP students have upgraded in this way. Despite some early concerns that the students might all wish to stay on in Cape Town they have in fact spread themselves among the partner institutions without any outside pressure.

NASSP graduates who do not continue in research go into industry or commerce – taking with them practical skills in problem solving, data analysis, computer programming and science communication that will serve them well in any challenging career. We thus anticipate that NASSP will make a broader contribution to the skills shortage in the country, and not simply provide more astronomers.

A steering committee, comprising representatives of the participating institutions, meets at least once a year and decides the broad policy for NASSP, including courses, lecturers and admissions policy; this is chaired by the author. The day to day running of NASSP has, since 2007, been overseen by an Executive Committee chaired by the NASSP Coordinator (normally Peter Dunsby who is currently on sabbatical) and comprising representatives of the main organizations or departments based in Cape Town. The members of this committee are listed in the Appendix.

## NASSP Students

Table 2 shows the numbers of students who have graduated from the NASSP honours programme since it started in 2003, and their demographics. The numbers who failed or dropped out are small, averaging one per year. To put these numbers in perspective one needs to appreciate that there had been no honours graduates in astronomy in the two years prior to NASSP and more importantly no students had started a masters degree in astronomy. Viewed against this background NASSP is highly successful, but there are many challenges, some of which are described below.

**Table 2: NASSP Honours Students**

| Year | Male | Female | Black | White | total |
|---|---|---|---|---|---|
| 2003 | 7 | 5 | 5 | 7 | **12** |
| 2004 | 7 | 3 | 7 | 3 | **10** |
| 2005 | 8 | 3 | 7 | 4 | **11** |
| 2006 | 14 | 2 | 8 | 8 | **16** |
| **total** | **36** | **13** | **27** | **22** | **49** |

At the masters level the numbers are similar, but smaller. They are less easy to interpret in that several of students have not graduated because they are taking longer than a year, sometimes very much longer, to complete their thesis. In many cases this has more to do with the complexity of the research topic than the ability of the student. Rectifying this situation is one of NASSP's challenges.

The students all receive grants which cover their fees and basic cost of living. The grants are generous by South African standards, although they would look small to Americans. Being able to offer financial assistance to students has been a critical factor in NASSP's success and has only been possible because of generous donations from the Ford Foundation, the NRF, the Canon Collins Educational Trust, the UCT Vice Chancellor's Fund and the Mellon Foundation. In 2006 NASSP received direct support from the DST which will ensure it has a future, provided certain conditions are met.

The Southern African Large Telescope (SALT) was envisaged as a facility for the region, not simply for South Africa. MeerKAT will probably have outstations in neighbouring countries, and if South Africa wins the bid to host the SKA its remote antennae will spread over much of the continent. For these and other reasons NASSP, from the outset, aimed to recruit from Africa, not simply South Africa. To date students have come from Botswana, Ethiopia, Gabon, Kenya, Madagascar, Mozambique, Rwanda, Sudan, Uganda, Zambia and Zimbabwe; often successive generations from the same home institution. These students come well prepared scientifically (better than many South African graduates), they are strongly motivated and work hard. So far all of the students recruited from outside of the country have graduated. The most serious challenge for many of them is the English language. Extra

English tuition is now a feature of the first semester of the honours course for anyone who needs it, including South Africans.

The most serious challenge to NASSP itself is recruiting and graduating black South Africans. In 2003, 2004 and 2005 there were 2 per year and in 2006 only 1 black South African honours graduate. The continuation of NASSP's funding from the government depends on increasing the numbers. The reasons for the paucity of local black students are multifaceted and linked to the problem mentioned above, that very few black students qualify to study physical science at university at all. Those who qualify and are good can demand very high salaries without postgraduate qualifications; when your family is depending on you the pressure to get a well paid job immediately is huge. Other factors, including low salaries and a lack of black role-models, contribute to the low enrolment numbers.

NASSP is fortunate that towards the end of 2006 the University of the Western Cape (UWC) agreed to join the collaboration. UWC is an historically black university with a record for producing competent black graduates. Their experience should prove invaluable and the recruitment of astronomers to their academic staff will allow for their full participation in NASSP from 2007.

NASSP has been reasonably successful at recruiting women students. They have frequently been top of the honours classes and many of them have gone on to do PhDs.

## *NSBP participation in NASSP*

Shortly after the meeting in Boston at which this paper was presented we heard that NSBP had received a grant from the Kellogg foundation that will facilitate the participation of Americans, particularly African Americans, in NASSP. The programme will be coordinated by Charles McGruder (charles.mcgruder@wku.edu) and it will enable US Faculty to visit Cape Town for the purpose of giving lecture courses at NASSP; research collaborations with South Africa will also be facilitated. There will also be grants for two USA students to study at NASSP. Anyone interested should contact Prof McGruder.

## ACKNOWLEDGMENTS

I am very grateful to NSBP and particularly to Lawrence Norris and Charles McGruder for making it possible for me to participate in this meeting of NSBP and NSHP.

# REFERENCES

1. Department of Science and Technology, *South Africa's National Research and Development Strategy*, Pretoria, SA Government, 2002.
2. P. A. Whitelock, "Optical Astronomy in Post-Apartheid South Africa: 1994 to 2004" in *Organizations and Strategies in Astronomy Vol 5*, edited by A. Heck, Dordrecht, Kluwer, 2004, pp. 30-60.

# APPENDIX - NASSP EXECUTIVE COMMITTEE

- Prof Peter Dunsby (Coordinator), Department of Mathematics and Applied Mathematics, UCT (peter.dunsby@uct.ac.za).
- Prof Renée Kraan-Korteweg, Department of Astronomy, UCT (kraan@circinus.ast.uct.ac.za).
- Dr Thebe Medupe, Department of Astronomy, UCT & SAAO (thebe@saao.ac.za).
- Prof Daniel Adams, Department of Physics, University of the Western Cape (dadams@uwc.ac.za).
- Prof Patricia Whitelock, Department of Mathematics and Applied Mathematics and Department of Astronomy, UCT, and SAAO.
- Ms Penny Middlekoop, NASSP Administrator, UCT (Penny.Middlekoop@uct.ac.za).